\documentclass[aps,prd, twocolumn,superscriptaddress,nofootinbib,preprintnumbers]{revtex4-1}

\usepackage{hyperref} % should be commented out when submitting to arXiv!!!

\usepackage{amsmath,amsfonts,amssymb}
\usepackage{color,slashed,mathrsfs}
\usepackage{epsfig,psfrag}

\def\iab{\mathrm{ab}^{-1}} % fb^-1
\def\TeV{\mathrm{TeV}}     % TeV
\def\GeV{\mathrm{GeV}}     % GeV
\def\MeV{\mathrm{MeV}}     % MeV
\def\missET{\slashed E_\mathrm{T}} % missing E_T
\def\O{\mathcal{O}}

\allowdisplaybreaks

\makeatother

\allowdisplaybreaks

\begin{document}

\title{Dark Matter Search at Colliders and Neutrino Floor}

\author{Qing-Hong Cao}
\email{qinghongcao@pku.edu.cn}
\affiliation{Department of Physics and State Key Laboratory of Nuclear Physics and Technology, Peking University, Beijing 100871, China}
\affiliation{Center for High Energy Physics, Peking University, Beijing 100871, China}
\affiliation{Collaborative Innovation Center of Quantum Matter, Beijing 100871, China}

\author{An-Kang Wei}
\email{ankangwei@pku.edu.cn}
\affiliation{Department of Physics and State Key Laboratory of Nuclear Physics and Technology, Peking University, Beijing 100871, China}

\author{Qian-Fei Xiang}
\email{xiangqf@pku.edu.cn}
\affiliation{Center for High Energy Physics, Peking University, Beijing 100871, China}

\begin{abstract}
The sensitivity of direct detection of dark matter (DM) approaches the so-called neutrino floor below which it is hard to disentangle the DM candidate from the background neutrino. In this work we consider the scenario that no DM signals are reported in various DM direct detection experiments and explore whether the collider searches could probe the DM under the neutrino floor.  We adopt several simplified models in which the DM candidate couples only to electroweak gauge bosons or leptons in the standard model through high dimensional operators. After including the RGE running effect we investigate constraints from direct detection, indirect detection and collider searches. The collider search can probe a light DM below neutrino floor. Especially, for the effective interaction of $\bar{\chi}\chi B_{\mu\nu}B^{\mu\nu}$, current data of the mono-photon channel at the 13 TeV LHC has already covered entire parameter space of the neutrino floor. 
\end{abstract}
%
%\pacs{***, ***}
%
\maketitle

\section{Introduction}

The existence of dark matter (DM)  has been well established by numerous astrophysical and cosmological observations, especially the very precise measurement of the cosmic microwave background (CMB)~\cite{Hinshaw:2012aka,Aghanim:2018eyx}. The most popular and attractive DM candidates  are the so-called weakly interacting massive particles (WIMPs).
 Assuming standard thermal history of cosmology,  the WIMPs produced in the early Universe through thermal freeze-out naturally  give the  observed DM relic density (``WIMP miracle")~\cite{Jungman:1995df}.  Because of their weak interactions with Standard Model (SM) particles, the WIMPs with mass around $100~{\rm GeV}$ would create significant signals in direct and indirect detection experiments, and they could also be copiously produced at high-energy colliders. 
In the last two decades the precision of DM direct detection experiments has been improved significantly and is approaching the  ``neutrino floor" which is an intrinsic background of the DM direct detection~\cite{Freese:2012xd,Ahlen:2009ev,Grothaus:2014hja,Billard:2013qya,Ruppin:2014bra}. 
However, till now, null results are reported in all kinds of dark matter searching experiments, which challenge the WIMP assumptions of the DM.
As demonstrated in Ref.~\cite{Cheung:2012gi},  if the DM candidate couples to quarks directly,  the parameter regions to give right relic density are not consistent with the null results of  direct detection, indirect detection, and collider searches. That motivates us to consider about the possibility that the DM candidate interacts only with electroweak gauge bosons~\cite{Cirelli:2005uq,Cirelli:2009uv,Cai:2012kt,Cai:2015kpa,Cohen:2011ec, Fischer:2013hwa,  Dedes:2014hga, Fedderke:2015txa, Calibbi:2015nha, Yaguna:2015mva, Tait:2016qbg, Horiuchi:2016tqw, Cai:2016sjz, Abe:2017glm, Cai:2017wdu, Xiang:2017yfs, Wang:2017sxx,Cao:2018nbr} or leptons in the SM~\cite{Bergstrom:2008gr,Barger:2008su,Cirelli:2008pk,Yin:2008bs,Zhang:2008tb,Fox:2008kb,Bergstrom:2009fa,Cao:2009yy,Ibarra:2009bm,Lin:2014vja,Xiang:2017jou}. 
 
Rather than focusing on specific DM models, we use an effective field theory (EFT) approach to parametrize the interactions of the DM candidate and the SM particles at new physics scale $\Lambda$. 
More specifically,  the interactions with gauge bosons are described by dimension-7 operators, while the interactions with leptons are described by dimension-6 operators. The DM candidate is considered as the only new particle in the energy relevant for current experiments, and those high-dimensional effective operators are presumably generated by new heavy particles of the dark sector which are much heavier than the DM candidate. Even though the DM candidate does not directly interact with quarks in the SM, the quantum effects can induce such interactions and yield a direct detection signal.
Hence one important question is that, when DM direct detection experiments reach the neutrino floor, whether one can use other experiments to explore the properties of DM.

One important feature is that energy scales involving in different experiments are different.
For instance, the momentum  exchange involved in DM-nucleus recoil is ~$\mathcal{O}(100)~\MeV$, while it is $\mathcal{O}(100)~\GeV$ in the LHC searches.
Different phenomena are created by operators generated at different energy scales; These operators themselves are related by renormalization group equations (RGE).
These loop effects or RGE effects have been widely studied assuming DM interacts with gauge bosons~\cite{Weiner:2012cb, Frandsen:2012db,Crivellin:2014gpa} and with leptons~\cite{Kopp:2009et,Fox:2011fx, DEramo:2016gos,DEramo:2014nmf}.

In this work, we evolve the RGE  from the scale $\Lambda$ down to the direct detection scale $\mu_D $.
Physical observations, calculated at corresponding energy scale, are compared with experimental observations.
We show that the collider searches can explore some regions under the neutrino floor.

The rest of the paper is organized as follows.
In Sec.~\ref{sec:eff_ww}  we give a brief description of DM interactions with electroweak gauge bosons, identify the parameter regions that could over close the universe, and that could be explored by  DM direct detection experiments and LHC searches.
In Sec.~\ref{sec:eff_lep}, we repeat the calculations, but for DM interacts with leptons, in that case  we also identify parameter regions that could be explored by future electron-positron colliders and DM indirect detection experiments.
Finally, we conclude in Sec.~\ref{sec:con}.

\section{Dark Matter candidate couples to gauge boson}
\label{sec:eff_ww}

We start with the case that the DM candidate is a Dirac fermion ($\chi$) and it interacts only with the electroweak gauge bosons in the SM. We first analyze the effective operators that could contribute to the DM direct detection when gradually evolving from the new physics (NP) scale $\Lambda$ down to the scale $\mu_D$, the characteristic energy scale of the DM direct detection.  

\subsection{Operator analysis}

At the level of dimension-5, there are only two operators as follows: 
\begin{equation}
\bar{\chi} \gamma^{\mu\nu}\chi B_{\mu\nu}~~~~~ \mathrm{and}  ~~~~~\bar{\chi} \gamma^{\mu\nu}\chi \tilde{B}_{\mu\nu},
\end{equation}
where $\gamma^{\mu\nu}\equiv [\gamma^\mu,\gamma^\nu]/4$ and $B_{\mu\nu}\equiv\partial_\mu B_\nu-\partial_\nu B_\mu$  are the field strength tensors of the $U(1)_Y$ gauge group. 
These two operators correspond to the weak magnetic dipole and electric dipole of DM, respectively, and induce unsuppressed cross sections of the DM annihilation into the $\gamma\gamma$ and $\gamma Z$ modes,  yielding a significant line spectrum of cosmic gamma-ray and thus are tightly constrained by current indirect detection experiments~\cite{Ackermann:2015lka}.
At the level of dimension-7,  there are four scalar-type operators built from $\bar{\chi}\chi$ or $\bar{\chi} \gamma^5 \chi$ as follows:
\begin{align}
 &\bar{\chi}\chi  B_{\mu\nu} B^{\mu\nu} , && \bar{\chi}\chi  W_{\mu\nu}^i W^{i\mu\nu} ,  \nonumber\\
 &\bar{\chi} \gamma^5 \chi  B_{\mu\nu}  \tilde{B}^{\mu\nu} , &&\bar{\chi} \gamma^5 \chi  W_{\mu\nu}^i \tilde{W}^{i\mu\nu},
 \label{eq:dim7}
\end{align}
and  six tensor operators constructed from $\bar{\chi}\gamma^{\mu\nu}\chi$, which are
\begin{equation}
\bar{\chi}\gamma^{\mu\nu}\chi B_{\alpha\mu} \tilde{B}^{\alpha\nu},~~~~~ \bar{\chi}\gamma^{\mu\nu}\chi W_{\alpha\mu}^i \tilde{W}^{i\alpha\nu}
\label{eq:ccbb1}
\end{equation}
and 
\begin{align}
&\bar{\chi}\gamma^{\mu\nu}\chi  B_{\mu\nu} |\Phi|^2,  && \bar{\chi}\gamma^{\mu\nu}\chi  \tilde{B}_{\mu\nu} |\Phi|^2,\nonumber\\
&\bar{\chi}\gamma^{\mu\nu}\chi  W_{i\mu\nu} \Phi^\dagger \tau^i \Phi, && \bar{\chi}\gamma^{\mu\nu}\chi  \tilde{W}_{i\mu\nu} \Phi^\dagger \tau^i \Phi.
\label{eq:ccbb2}
\end{align}
Here, $W^i_{\mu\nu}\equiv \partial_\mu W_\nu^i-\partial_\nu W_\mu^i+g_2\epsilon^{ijk}W_\mu^jW_\nu^k$  is the field strength tensor of the $SU(2)_L$ gauge group and $\tilde{W}^{i\mu\nu}$ is the corresponding dual tensor.
The last two operators in Eq.~\ref{eq:dim7} and the two operators in Eq.~\ref{eq:ccbb1} lead to unsuppressed $\gamma\gamma$ and $\gamma Z$ signals, and the four operators in Eq.~\ref{eq:ccbb2} induce unsuppressed $\gamma h$ signals; therefore, they are highly constrained and are ignored hereafter. For more discussions and phenomena about these operators, please refer to Ref.~\cite{Rajaraman:2012db, Frandsen:2012db,Crivellin:2014gpa}.
Finally, we end up with only the first two operators in Eq.~\ref{eq:dim7} and the Lagrangian can be expressed as 
\begin{equation}
 \label{eq:lan_g}
 \mathcal{L}_{eff} = \frac{C_B}{\Lambda^3} \bar{\chi}\chi B_{\mu\nu}B^{\mu\nu} +   \frac{C_W}{\Lambda^3}  \bar{\chi}\chi W_{\mu\nu}^iW^{i\mu\nu},
 \end{equation}
which yield velocity suppressed annihilation cross sections and therefore are free from the tight constraint of the gamma ray line spectrum observations~\cite{Ackermann:2015lka}. $C_{B,W}$ denotes the Wilson coefficient of the corresponding operator.  

We assume only these two operators are generated at the scale $\Lambda$ when other new physics resonances much heavier than $\Lambda$ are all decoupled. Even though the DM candidate $\chi$ and the SM quarks are not directly coupled at the NP scale $\Lambda$, they are linked by quantum effects at a lower scale through the RGE running.  
For example, as shown in Ref.~\cite{Crivellin:2014gpa}, two extra operators coupling the DM candidate to quarks, i.e. 
\begin{equation}
\O_y=y_q\bar{\chi}\chi\bar{q}\phi q ~~~\mathrm{and}~~~  \O_{\phi}=\bar{\chi}\chi(\phi^\dagger\phi)^2,
\end{equation}
are generated when evolving from $\Lambda$ down to the weak scale. 
Here, $y_q=\sqrt{2}m_q/v$ is the Yukawa coupling and $v$ is the vacuum expectation value (VEV) of the SM Higgs doublet $\phi$.
It is straightforward to show that, to the leading logarithmic (LL) order,  the Wilson coefficients of the operator of $\O_{y,\phi}$ and $\O_{B,W}$ are related as follows:
\begin{eqnarray}
C_y^q({\mu}) & \simeq & \frac{3 Y_{qL} Y_{qR} \alpha_1}{\pi} \mathrm{ln}\left(\frac{\mu^2}{\Lambda^2}\right) C_B(\Lambda), \\
C_\phi({\mu} )& \simeq &- \frac{9 \alpha_1^2}{2} \mathrm{ln}\left(\frac{\mu^2}{\Lambda^2}\right) C_W(\Lambda),
\end{eqnarray}
where $Y_{qL} $ ($ Y_{qR}$) are the hypercharges of the left-handed (right-handed) quarks assigned as $Y_{uL} = Y_{dL} = 1/6$, $Y_{uR}=2/3$ and $Y_{dR} = -1/3$.
$\alpha_1$ and $\alpha_2$ are the gauge coupling constants of $U(1)_Y$ and $SU(2)_L$ gauge group, respectively.
At the weak scale $\mu_Z(\equiv m_Z)$, $\alpha_1(\mu_Z) \simeq 1/98$ and $\alpha_2(\mu_Z) \simeq 1/29$.

After electroweak symmetry-breaking (EWSB), $B_\mu$ and $W_\mu^3$ mix into the photon field $A_\mu$ and massive gauge field $Z_\mu$.
The effect of $Z_\mu$ decouples at lower  scale below $\mu_Z$ and the relevant  operator basis become
\begin{equation}
\O_A=  \bar{\chi}\chi F_{\mu\nu}F^{\mu\nu}, \qquad \O_q = m_q  \bar{\chi}\chi \bar{q} q.
\label{eq:OAq}
\end{equation}
The matching conditions between these two operator bases are
\begin{align}
\label{eq:CA}
&C_A(\mu_Z)=  c_W^2 C_B(\mu_Z) + s_W^2 C_W(\mu_Z), \nonumber\\
& C_q(\mu_Z)=   C_y^q(\mu_Z) - \frac{v^2}{m_h^2} C_\phi(\mu_Z),
\end{align}
where $c_W \equiv \mathrm{cos}\theta_W$ and $s_W \equiv \mathrm{sin}\theta_W$ are related to the Weinberg angle $\theta_W$.

Next, we further evolve RGEs from $\mu_Z$ down to the hadronic scale $\mu_{D} \sim 1~\GeV$ at which the DM candidate interacts with the nucleons. $\O_q$ will get contributions from $\O_A$ through exchanging of virtual photons~\cite{Frandsen:2012db}.
To LL accuracy, 
\begin{equation}
C_q(\mu) \simeq C_q(\mu_Z) + \frac{3 Q_q^2 \alpha}{\pi} \mathrm{ln}(\frac{\mu^2}{\mu_Z^2}) C_A(\mu_Z)
\label{eq:cq}
\end{equation}
for $m_q < \mu < \mu_Z$ with $Q_q$ is the electric charge of quarks.
During the evolution of RGEs, heavy quarks in the SM are integrated out, yielding an operator
\begin{equation}
\O_G =  \alpha_s \bar{\chi}\chi G_{\mu\nu}^a G^{a,\mu\nu},
\label{eq:OG}
\end{equation}
where $G_{\mu\nu}^a$ denotes the field strength of gluon. 
The matching condition is given by the simple replacement~\cite{Shifman:1978zn}
\begin{equation}
C_t m_t \bar{\chi}\chi \bar{t} t \to C_G  \alpha_s \bar{\chi}\chi G_{\mu\nu}^a G^{a,\mu\nu},
\end{equation}
with $C_G$ given at the leading order by 
\begin{equation}
 C_G(m_t)  = -\frac{1}{12\pi} C_t (m_t).
\end{equation}

\begin{figure}
\includegraphics[width=.38\textwidth]{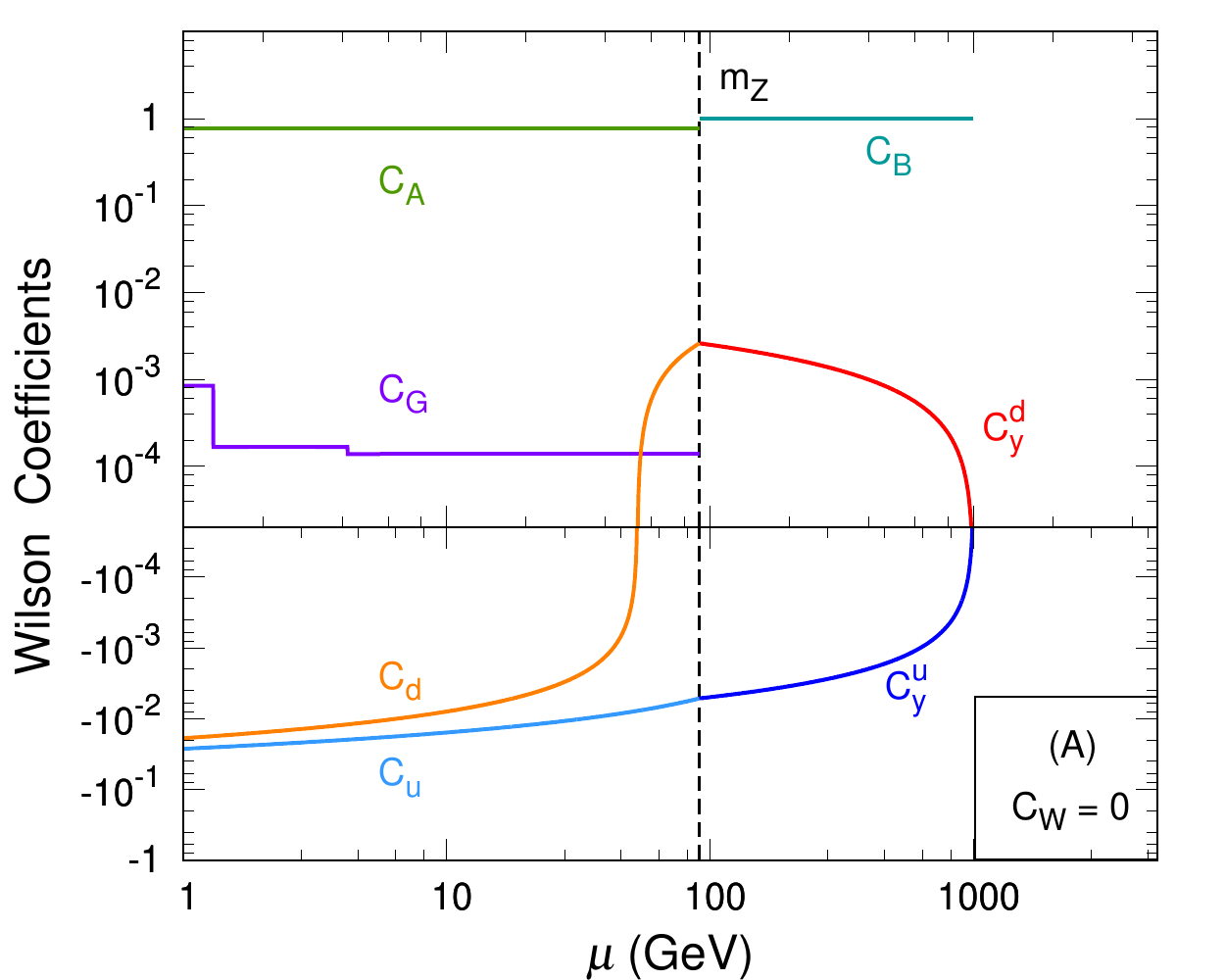}
\includegraphics[width=.38\textwidth]{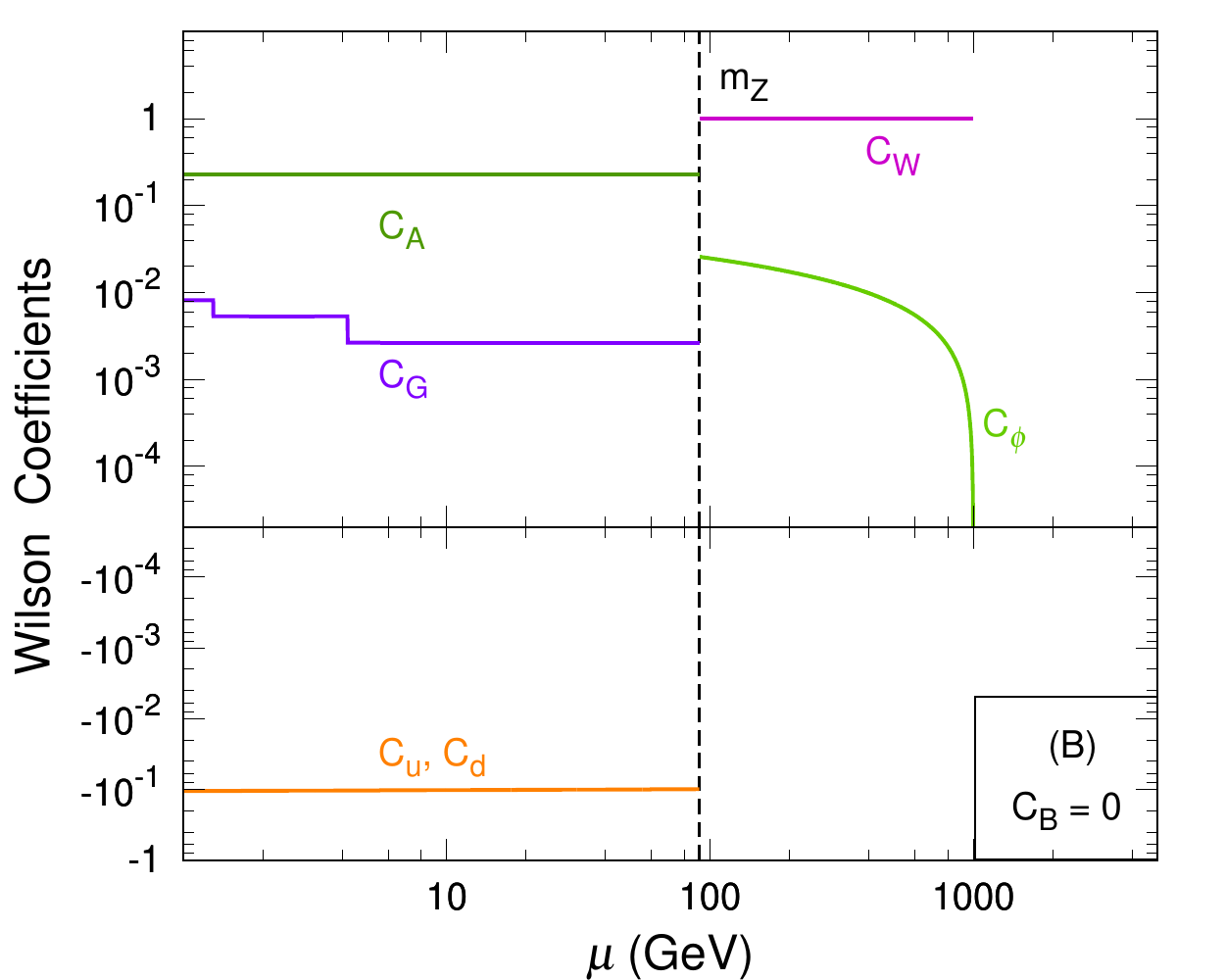}
\caption{Illustrations of running of Wilson coefficients for $C_W=0$ (A) and $C_B=0$ (B).}
\label{fig:rge_ww}
\end{figure}

Finally, we end up with the operator basis consisting of $\O_q$, $\O_A$, and $\O_G$.
The mixing of $\O_G$  with  $\O_q$ and $\O_A$ are subdominant and can be safely ignored.
Taking into account of the threshold effects at $m_b$ and $m_c$ which contribute to $\mathcal{O}_G$,
we obtain the final expressions for the Wilson coefficients as follows:
 \begin{align}
 C_q(\mu)&\simeq\bigg(\frac{3Y_{qL}Y_{qR}\alpha_1}{\pi}C_B(\Lambda)+\frac{9\alpha_2^2}{2}\frac{v^2}{m_h^2}C_W(\Lambda)\bigg)\mathrm{ln}\bigg(\frac{m_Z^2}{\Lambda^2}\bigg)\nonumber \\
 &+\frac{3Q_q^2\alpha}{\pi}C_A(\mu_Z)\mathrm{ln}\bigg(\frac{\mu_D^2}{m_Z^2}\bigg),\nonumber\\
 C_G(\mu)&\simeq-\frac{1}{12\pi}\bigg\{\bigg(\frac{\alpha_1}{2\pi}C_B(\Lambda)+\frac{27\alpha_2^2}{2}\frac{v^2}{m_h^2}C_W(\Lambda)\bigg)\mathrm{ln}\bigg(\frac{m_Z^2}{\Lambda^2}\bigg)\nonumber\\
 &+\frac{\alpha}{3\pi}C_A(\mu_Z)\bigg[\mathrm{ln}\bigg(\frac{m_b^2}{m_Z^2}\bigg)+4\mathrm{ln}\bigg(\frac{m_c^2}{m_Z^2}\bigg)\bigg]\bigg\}.
  \label{eq:Wc}
 \end{align}
Figure~\ref{fig:rge_ww} displays the RGE running of Wilson coefficients. For simplicity, we consider one operator at a time,  i.e.,  either $C_W=0$ (A) or $C_B=0$ (B), and fix the cut off scale $\Lambda$ to be $1000\ \mathrm{GeV}$. 
In the case of $C_W=0$, the Willson coefficients involving up-type quark ($C_y^u$) and down-type  quark  ($C_y^d$) exhibit opposite sign as the hypercharges of up-type quark and down-type quark are different; see the red and blue curves in the top figure. The sign difference remains unchanged even after the EWSB and matching. However, running from the weak scale $\mu_Z$ down to the hadronic scale $\mu_D$, the correction of $C_A$ to $C_d$ is significant such that it changes the sign of $C_d$ from positive to negative. On the other hand, in the case of $C_B=0$, both the $C_u$ and $C_d$ coefficients receive identical contributions from $C_\phi$ while the contributions from $C_A$ can be safely ignored; see the bottom figure.

\subsection{Experimental searches}

As an important approach to probe the DM candidate, the direct detection experiments search for recoil signals of target nuclei scattered off by incident DM particles.
Even though the DM candidate in our model does not interact with quarks directly at the scale $\Lambda$, the direct detection signal could be induced at the hadron scale. 
For example, quantum loop effects induce the interactions of the DM candidate with quarks and gluons, e.g., $\O_q$ and $\O_G$ in Eq.~\ref{eq:OAq} and Eq.~\ref{eq:OG}, which result in the interaction between the DM candidate and nucleon. 
In addition, two virtual photons exchange will induce interaction with the entire nucleus, with interaction strength proportional to the total electric charge $Ze$ of the nucleus~\cite{Weiner:2012cb}.

The contributions of these operators are coherent, leading to the spin-independent (SI) cross section for DM scattering with nuclei~\cite{Weiner:2012cb,Frandsen:2012db,Crivellin:2014gpa},
\begin{equation}
 \sigma_N^{\mathrm{SI}}\simeq\frac{m_{\mathrm{red}}^2m_N^2}{\pi\Lambda^6}\bigg|\frac{\alpha Z^2}{A}f_F^N C_A(\mu_D)
 +\frac{Z}{A} f_p
 +\frac{A-Z}{A} f_n 
 \bigg|^2,
 \label{eq:SI}
\end{equation}
where $m_{\mathrm{red}}$ labels the reduced mass of DM-nuclei system and $m_N$ is the mass of  nucleon, $f_F^N$ is the form factor  for photon and equals to $0.08$ (0.12) for Xenon  (Argon) target.
The form factor $f_N$ ($N=p, n$) describes the interaction of the DM  candidate to nucleon, and they are related to the interactions with quarks through
\begin{equation}
 f_N  =\sum_{q=u,d,s}f_q^N C_q(\mu_D)-\frac{8\pi}{9}f_G^N C_G(\mu_D),
\end{equation}
where the nucleon form factors $f_q^N$'s are given by~\cite{Belanger:2013oya} 
\begin{align}
&f_d^p = 0.0191, &&f_u^p=0.0153, & f_d^n = 0.0273,\nonumber\\
&f_u^n=0.0110, & &f_s^p = f_s^n=0.0447, 
\end{align}
and 
\begin{align}
&f_G^N  = 1- \sum_{q=u,d,s} f_q^N.
\end{align}
The contributions of proton and neutron are separated in Eq.~\ref{eq:SI} as the DM candidate couples differently to proton and neutron.

\begin{figure}
\includegraphics[width=.4\textwidth]{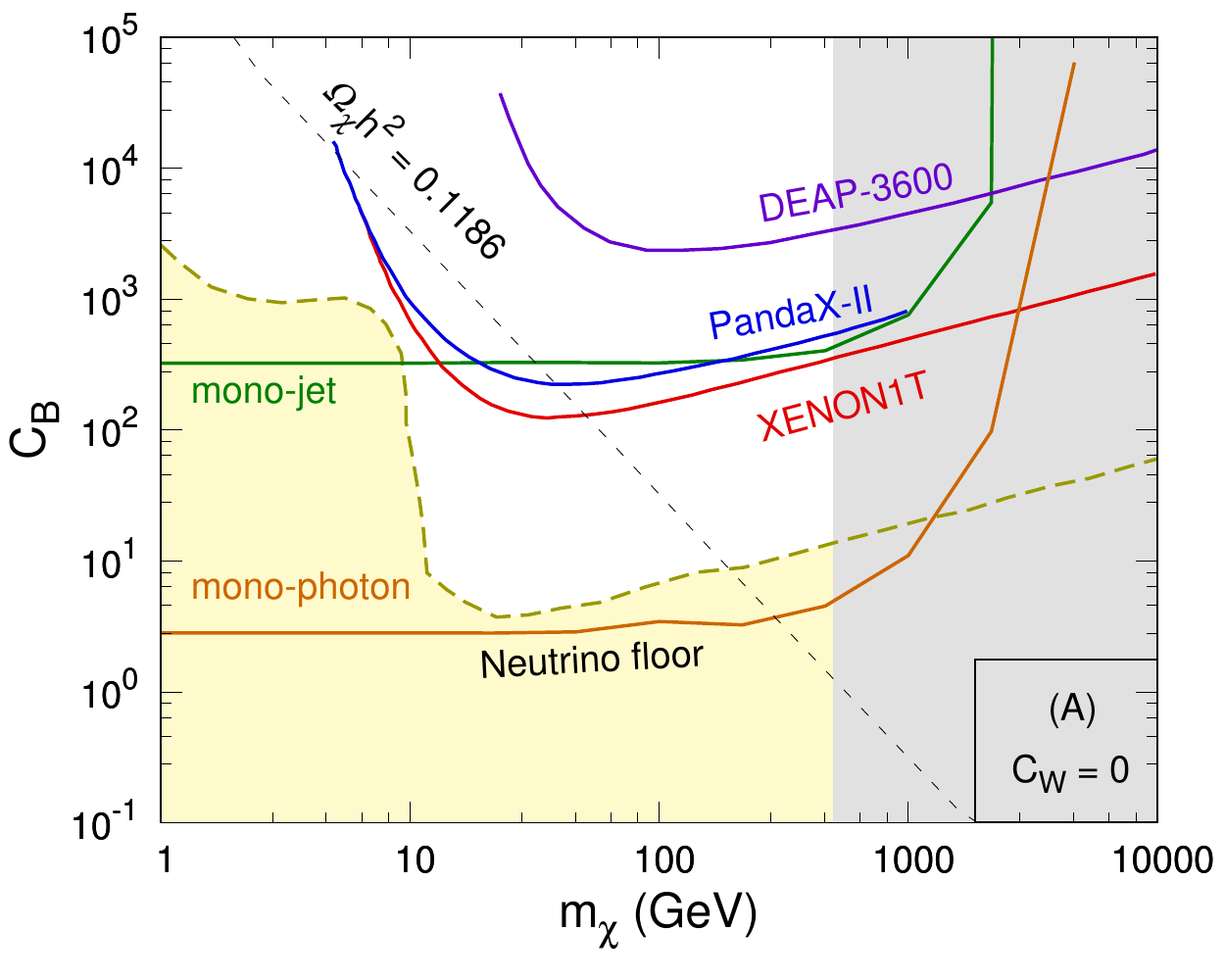}
\includegraphics[width=.4\textwidth]{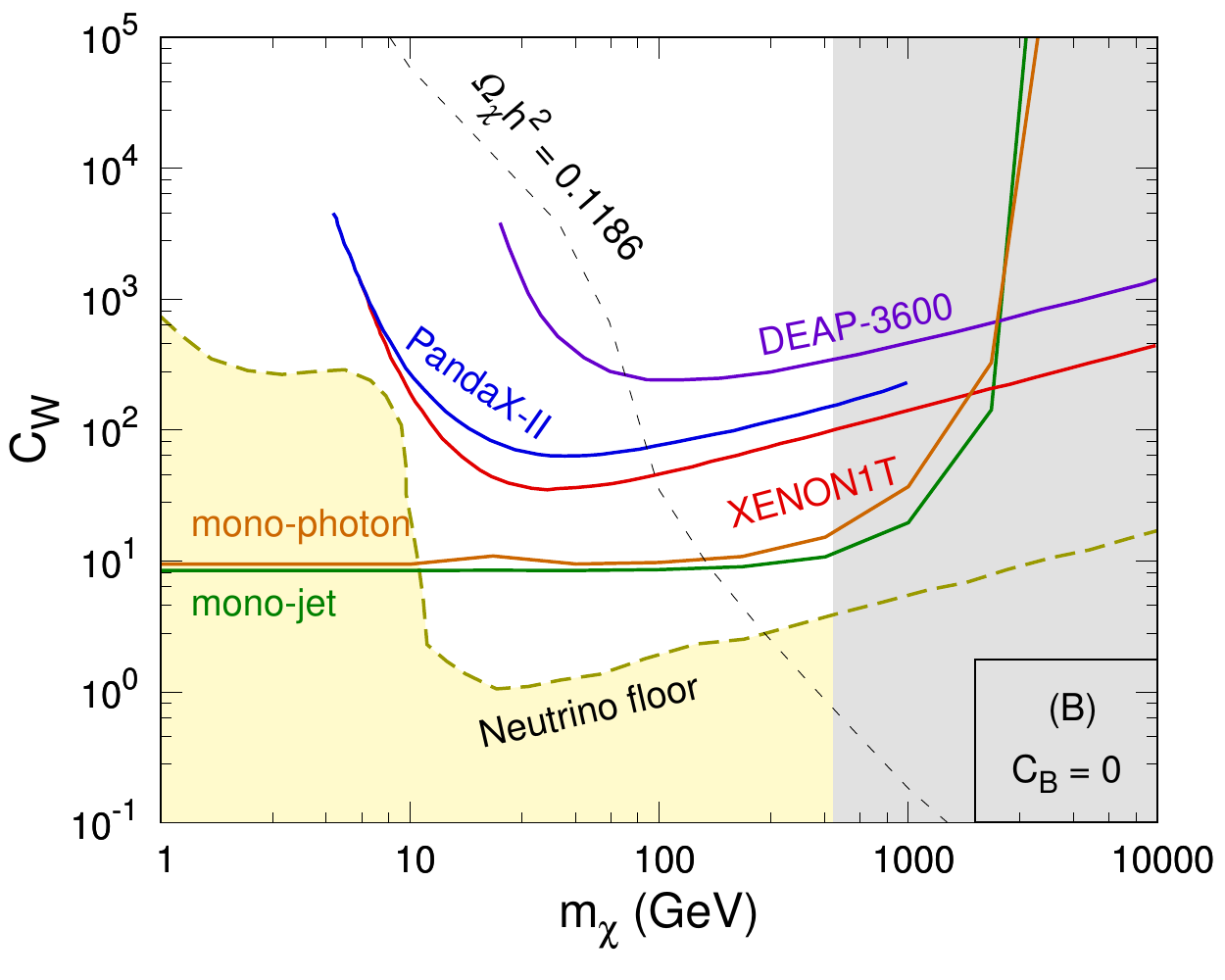}
\caption{Constraints in the $m_\chi$-$C_B$  plane (A) and $m_\chi$-$C_W$ plane (B) with $\Lambda=1~{\rm TeV}$.
Brown  and  green solid lines denote the exclusion limits from the mono-photon~\cite{Aaboud:2017dor} and mono-jet~\cite{Aaboud:2017phn} searches at the $95\%$ confidential level at the 13~TeV LHC, respectively.
For the SI DM-nucleon scattering, recent bounds from XENON1T~\cite{Aprile:2017iyp}, PandaX-II~\cite{Tan:2016zwf},  and DEAP-3600~\cite{Ajaj:2019imk} are shown.
The gray shaded region denotes the parameter space where the EFT is invalid, estimated as $m_\chi > \Lambda/2$. For illustration the contours of the relic abundance, $\Omega_\chi h^2=0.1186$, are also plotted; see the dark dashed curves. }
\label{fig:eff_ww}
\end{figure}

Figure~\ref{fig:eff_ww} displays the constraints on the Wilson coefficients as a function of $m_\chi$ at the $90\%$ confidence level (C.L.) from the PandaX-II~\cite{Tan:2016zwf}, XENON1T~\cite{Aprile:2017iyp}, and DEAP-3600~\cite{Ajaj:2019imk}.
For illustration we fix the cutoff scale as $\Lambda=1~\TeV$ and $C_W=0$ ($C_B=0$) in the  top (bottom) of Fig.~\ref{fig:eff_ww}, respectively. 
Both the PandaX-II and XENON1T experiments use the Xenon target, therefore, the contributions of $C_A$ are enhanced by a relatively large $Z^2$ and dominate over the contributions of $f_p$ and $f_n$. On the other hand, 
the contributions of $f_p$ and $f_n$ dominate in the DEAP-3600 experiment which uses the Argon target. 
For illustration we plot the contours of the correct relic abundance, $\Omega_\chi h^2 = 0.1186\pm 0.0020$~\cite{Aghanim:2018eyx}; see the Appendix for detailed calculations. 

The yellow regions in Fig.~\ref{fig:eff_ww} denote the so-called neutrino floor, representing the WIMP-discovery limit obtained by assuming an exposure of 1000 $^8\mathrm{B}$ neutrinos are detected on a Xenon target~\cite{Ruppin:2014bra}.
Recently, a few new methods have been proposed to improve the sensitivity of DM detection beneath the neutrino floor, e.g. 
using the annual modulation signal~\cite{Freese:2012xd}, directional detection methods~\cite{Ahlen:2009ev,Grothaus:2014hja}, or looking for a possible complementarity between different target nuclei~\cite{Ruppin:2014bra};
however, the power of these methods is still limited.  It is important to study whether the collider search 
can probe the DM candidate below the neutrino floor.

The processes of interests to us on the colliders are $q\bar{q}/gg\to \chi\bar{\chi}+j$ and $q\bar{q}/gg\to \chi\bar{\chi}+\gamma$, where $j$ denotes a light-flavor jet.  The DM candidate often appears as an invisible object on the colliders and yields a signature of missing transverse momentum ($\missET$). The event topology of the two signal processes consists of a large $\missET$ with either a hard jet or a hard photon. 
The former signature is often named as ``mono-jet" while the latter is called ``mono-photon". 
For the model considered in this section, operator $\O_A$ will induce $\text{mono-photon}+\missET$ signal, and operators $\O_q$ and $\O_G$ will induce $\text{mono-jet}+\missET$ signal. In order to get the sensitivities of LHC searches, we perform collider simulations of the mono-jet and mono-photon channels. The parton level events are generated by \texttt{MadGraph 5}~\cite{Alwall:2014hca}, and \texttt{PYTHIA~6}~\cite{Sjostrand:2006za} is used to deal with parton shower and hadronization. We adopt \texttt{Delphes~3}~\cite{deFavereau:2013fsa}  to carry out a fast detector simulation with a parameter setup for the ATLAS detector.

We follow the procedure of the ATLAS group of the $\text{mono-photon}+\missET$~\cite{Aaboud:2017dor} and $\text{mono-jet}+\missET$~\cite{Aaboud:2017phn} analysis with an integrated luminosity of $36.1~{\rm fb}^{-1}$ at the $\sqrt{s}=13~{\rm TeV}$ LHC.
Figure~\ref{fig:eff_ww} presents the exclusion limits derived from the mono-photon search (brown) and the mono-jet search (green) at the LHC.
On the one hand, it is hard to directly probe a DM candidate heavier than about $2~\TeV$ at the LHC as limited by the colliding energy. 
On the other hand, when the DM is light, say $m_\chi \lesssim 100~{\rm GeV}$, the limits are independent of the DM mass  as the production cross sections are mainly determined by the effective couplings.  
Obviously, the collider search has a better sensitivity than the direct detection experiments in the regime that the EFT is valid,  e.g. $m_\chi \lesssim 2~\TeV$. We emphasize that the collider searches can probe a light DM  ($m_\chi \lesssim 10~\GeV$) below neutrino floor.

In the case of $C_W=0$, the sensitivity of the mono-photon channel is much higher than the one of the mono-jet channel, cf. Fig.~\ref{fig:eff_ww}(A). It can be understood as follows. First, the mono-photon signal event can be generated by the $\O_A$ and $\O_q$, while the mono-jet signal event can be generated only by the $\O_q$. Note that the $\O_A$ is directly linked with the $\O_B$, while the $\O_q$ is generated through loop effects and is much smaller than the $\O_A$; see Fig.~\ref{fig:rge_ww}(A).  Second, the SM background of the mono-photon channel is much cleaner than those of the mono-jet channel, yielding a better sensitivity to the DM searches at colliders. As a result, the mono-photon channel can cover the entire parameter space of the neutrino floor. 

In the case of $C_B=0$, the $\O_A$ is suppressed by the weak mixing angle after EWSB. On the other hand , the $O_q$ is slightly enhanced after matching at the weak scale. As a result, the mono-photon and mono-jet channels yields comparable sensitivities to the DM searches; see Fig.~\ref{fig:eff_ww}(B). Even though both channels are better than the DM direct detection experiments, they cannot reach the neutrino floor for $m_\chi \gtrsim  10~{\rm GeV}$.

\section{Dark Matter candidate couples to leptons}
\label{sec:eff_lep}

In this section we study the scenario that a fermionic DM ($\chi$) interacts only with leptons at the NP scale $\Lambda$, named as lepton-philic DM. 
Assume the DM candidate $\chi$ interacts universally to all the leptons in the SM through
 \begin{equation}
 \mathcal{L}_{eff}^l = \frac{C_l}{\Lambda^2}  \sum_i \bar{\chi} \gamma_{\mu} \chi 
 (\overline{l_L^i}\gamma_\mu l_L^i + \overline{e_R^i}\gamma_\mu e_R^i),
 \end{equation}
 where the summation over the three generations of leptons in the SM is understood. Again, even though the DM candidate does not directly couple to the SM quarks at the scale $\Lambda$, the RGE running effects would induce non-zero interactions between the DM candidate and the quarks as follows: 
 \begin{align}
 \label{ob}
&\mathcal{O}_q^i =\bar\chi\gamma^\mu\chi\overline{q_L^i}\gamma_\mu q_L^i, \nonumber\\
&\mathcal{O}_u^i= \bar\chi\gamma^\mu\chi\overline{u_R^i}\gamma_\mu u_R^i,  \\
&\mathcal{O}_d^i =\bar\chi\gamma^\mu\chi\overline{d_R^i}\gamma_\mu d_R^i.\nonumber
 \end{align} 
At the first glance the DM candidate only sees the leptons but is blind to the quarks; however, the connection between the DM candidate and the SM quarks is built through the operator $ \bar\chi\Gamma^\mu\chi H^\dagger i\overleftrightarrow{D}_\mu H $~\cite{DEramo:2014nmf,DEramo:2016gos}.

The strategy of calculating the RGE running effects is the same as the one described in the  previous section.
 After matching, the interactions of interests to us are 
\begin{equation}
\mathcal{L} \subset \frac{C_u^V}{\Lambda^2} \bar\chi\gamma^\mu\chi \bar{u} \gamma_\mu u +  \frac{C_d^V}{\Lambda^2} \bar\chi\gamma^\mu\chi \bar{d} \gamma_\mu d +  \frac{C_e^V}{\Lambda^2}  \bar\chi\gamma^\mu\chi \bar{e} \gamma_\mu e.
\label{eq:xxqq}
\end{equation}
In principle both Yukawa interaction and gauge interaction would influence the running of RGEs.
However, in the case that the DM candidate interacts equally to the left-handed and right-handed leptons,  the contributions  of the Yukawa interactions cancel in each generation~\cite{DEramo:2016gos}. 
As a result, the induced interactions in Eq.~\ref{eq:xxqq} are independent of the quarks masses and yield identical Wilson coefficients for the three generations.
For the sake of simplicity, we ignore the index of generation hereafter. 

\begin{figure}
\includegraphics[width=.4\textwidth]{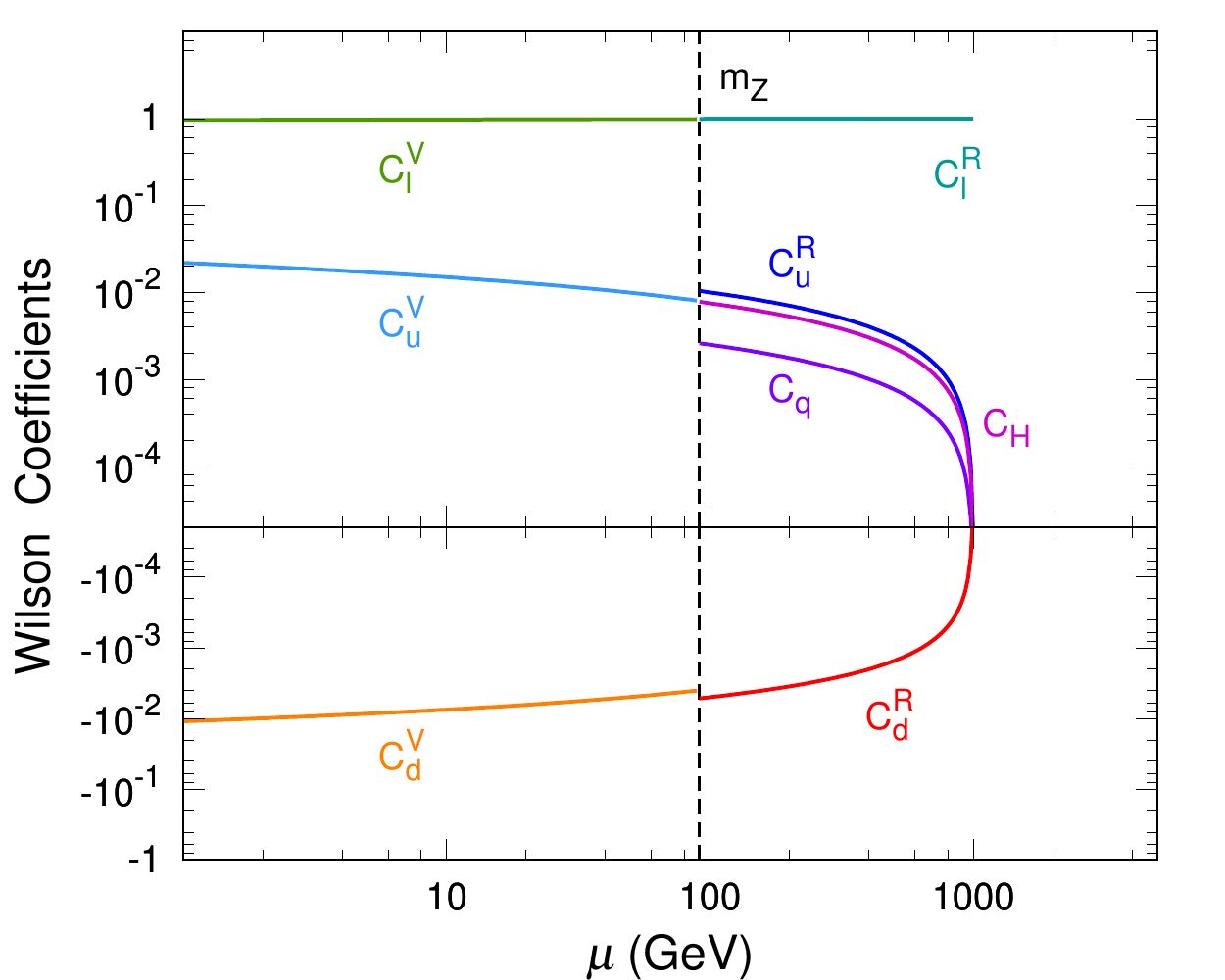}
\caption{ RGE running of Wilson coefficients of the effective couplings of $\chi$ to the SM fermions. }
\label{fig:rge_eff_lepton}
\end{figure}

We follow Refs.~\cite{DEramo:2014nmf,DEramo:2016gos}  and adopt the package \texttt{runDM}~\cite{runDM}  to perform a complete RGE running from the NP scale down to the scale of DM direct detections.
Figure~\ref{fig:rge_eff_lepton} displays the Willson coefficients for different operators.
The upper quarks and down quarks  pick up opposite sign with strength different by a factor of $2$ as the electromagnetic interaction plays the leading role in the RGE running~\cite{DEramo:2016gos}. 

Equipped with the effective couplings of the DM candidate to the SM quarks, we are ready to discuss the detection of the lepton-philic DM. 
First, consider the direct detection experiment. 
Owing to the conservation of vector current,  both sea quarks and gluons inside nuclei do not contribute. The contributions of all the valence quarks add coherently, leading to the WIMP-nucleus scattering cross sections as following:
\begin{equation}
 \sigma_N^{\mathrm{SI}} = \frac{m_{\mathrm{red}}^2}{\pi\Lambda^4}
 \bigg|
\frac{Z}{A} C_{p}^V+ \frac{A-Z}{A} C_{n}^V  
 \bigg|^2.
 \end{equation} 
Here, $C_{p}^V$ and $C_{n}^V $ denote the interactions of the DM candidate $\chi$ with proton and neutron, respectively. They are related to the effective couplings of $\chi$ to the $u$ quark and $d$ quark as follows:
\begin{align}
C_{p}^V = 2 C_u^V + C_d^V,\qquad C_{n}^V = C_u^V + 2 C_d^V,
\end{align}
where $C_u^V$  and $C_d^V$ can be approximated as~\cite{DEramo:2016gos}
 \begin{equation}
C_u^V \simeq  \frac{4\alpha}{3\pi} C_l,  \qquad C_d^V \simeq - \frac{2\alpha}{3\pi} C_l .
\end{equation}
Therefore, the interaction between $\chi$ and the neutron can be safely ignored.

 \begin{figure}
\includegraphics[width=.4\textwidth]{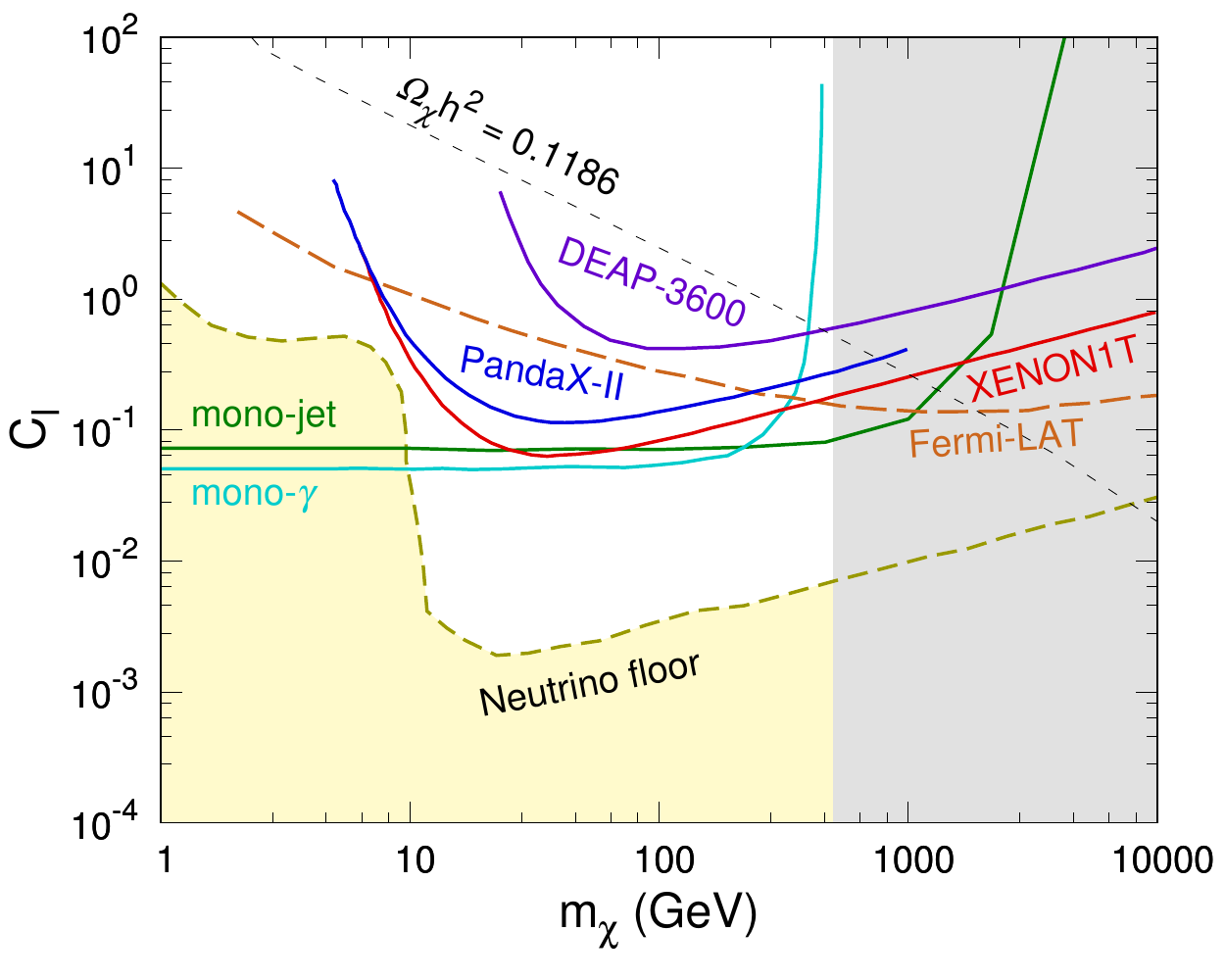}
\caption{The bounds in the $m_{\chi}$-$C_l$ plane with $\Lambda=1~{\rm TeV}$.
The brown dashed curve denotes the recent limit from the Fermi-LAT gamma-ray observations about the dwarf galaxies~\cite{Ackermann:2015zua}, and the cyan solid curve represents the projected sensitivity of the mono-$\gamma$ channel at the ILC.}
\label{fig:eff_lepton}
\end{figure}

Figure~\ref{fig:eff_lepton} shows the bounds on the Wilson coefficient $C_l$ at the $90\%$ confidence level obtained from various direct detection experiments.
The yellow shaded region denotes the parameter space below the neutrino floor. 
Similar to the case that DM candidate interacts only with electroweak gauge bosons,  the direct detection of the lepton-philic DM cannot touch the neutrino floor. 
Another way to search for the DM candidate is to examine high energy comic rays, gamma rays, and neutrinos induced by DM decays or annihilations in galactic and extragalactic objects. In the lepton-philic model the DM candidates predominantly annihilate into a pair of leptons. 
The constraint on $C_l$ from the indirect search of the DM through the cosmic gamma-ray observation by the Fermi-LAT collaboration~\cite{Ackermann:2015zua} is plotted in Fig.~\ref{fig:eff_lepton}; see the brown dashed curve. 
Even though the DM  will annihilate into all the three generation leptons, the $\tau^+\tau^-$ channel dominates in the gamma-ray observations as more photons are produced from the $\tau$-lepton decay.  Unfortunately, the indirect search experiment cannot reach the neutrino floor neither.

The effective couplings of $\chi$ to the SM quarks, given in Eq.~\ref{ob}, give rise to the mono-jet signal at the LHC; therefore, it yields a strong bound on $C_l$. See the green curve in Fig.~\ref{fig:eff_lepton}. 
It is obvious that LHC will constrain the DM $\chi$ with a mass up to about 2 TeV.
For $m_\chi \lesssim 10~\GeV$, the LHC can probe the parameter space below the neutrino floor. 

The lepton-philic DM can be better probed at future electron-positron colliders, e.g. CEPC~\cite{CEPCStudyGroup:2018ghi} and ILC~\cite{Baer:2013cma}. 
In this work we explore the potential of the ILC with $\sqrt{s}=1~\TeV$ and an integrated luminosity of $1~\iab$.
In order to identify the mono-$\gamma$ signal, we require that the signal event contains one energetic photon with energy $E_\gamma >  10~\GeV$  and  $10^\circ < \theta_\gamma < 170^\circ$, where $\theta_\gamma$ is the polar angle between the hard photon and the beam axis. 
We also require missing mass  $m_\mathrm{miss}  > 200~\GeV$, where the missing mass  $m_\mathrm{miss}$ is defined as $m_\mathrm{miss} = \sqrt{(p_{e^+}+ p_{e^-} + p_\gamma)^2}$ with  $p_{e^+}$ ($p_{e^-}$) is the 4-momentum of the initial positron (electron) and  $ p_\gamma$ is the 4-momentum of final photon~\cite{Yu:2013aca}.
The cyan solid line in Fig.~\ref{fig:eff_lepton} represents the projected sensitivity of the ILC experiment on $C_l$. 
The ILC has a better sensitivity than the LHC for the DM candidate in the mass region of interests to us, say $m_\chi \lesssim 200~\mathrm{GeV} $;
unfortunately, the collider searches cannot reach the neutrino floor for $m_\chi\gtrsim 10~{\rm GeV}$.

\section{Conclusions}
\label{sec:con}

With increasing exposures the sensitivity of direct detection of dark matter (DM) candidate approaches the so-called neutrino floor, below which it is hard to distinguish the signal induced either by a DM candidate or by a neutrino. In this study we consider a nightmare scenario that all the direct detection experiments report null results even though the detection sensitivity reaches the neutrino floor. We demonstrate that the collider searches can probe the parameter space under the neutrino floor.

For illustration, we adopt two simplified models in which the DM candidate only couples to electroweak gauge bosons or leptons. Rather than focusing on a specific theory model, we use an effective Lagrangian approach to parametrize the interactions of DM with the gauge bosons or the leptons in the SM at new physics scale $\Lambda$, assuming all  other heavy resonances in the UV complete model decouple at the scale $\Lambda$.  
Specifically, we consider two effective Lagrangians as follows:
\begin{align*}
 \mathcal{L}^{V}_{\rm eff} &= \frac{C_B}{\Lambda^3} \bar{\chi}\chi B_{\mu\nu}B^{\mu\nu} +   \frac{C_W}{\Lambda^3}  \bar{\chi}\chi W_{\mu\nu}^iW^{i\mu\nu},\nonumber\\
 \mathcal{L}_{\rm eff}^L &= \frac{C_l}{\Lambda^2}  \sum_i \bar{\chi} \gamma_{\mu} \chi 
 (\overline{l_L^i}\gamma_\mu l_L^i + \overline{e_R^i}\gamma_\mu e_R^i).
\end{align*}
Even though the DM candidate does not couple to the SM quarks directly at the scale $\Lambda$, it can interact with the quarks and gluons in the SM  through the RGE running effects. In this work we evolve the RGE from high scale $\Lambda$ to the EWSB scale, match to the basis of mass eigenstates, and then further evolve the RGE down to direct detection energy scale. For simplicity we consider one parameter at a time and set the other two coefficients to be zero.

We investigate relevant constraints from the direct/indirect detections and collider searches,  and find that the collider search has a better sensitivity than the direct and indirect detection experiments in the regime that the EFT is valid,  e.g. $m_\chi \lesssim 2~\TeV$.  In all the three simplified models the collider searches can probe a light DM  ($m_\chi \lesssim 10~\GeV$) below the neutrino floor.
More interestingly, in the case of $C_B \neq 0$ and $C_W=0$, the current data of the mono-photon channel at the 13 TeV LHC has already covered the entire parameter space of the neutrino floor.

\noindent{\bf Acknowledgments.~}
The work is supported in part by the National Science Foundation of China under Grant Nos. 11725520, 11675002, 11635001 and in part by the China Postdoctoral Science Foundation under Grant No. 8206300015.

\appendix
\section{DM annihilation and relic density}
\label{app:anni}

The relic abundance predicted by this model should be smaller than the observed one reported by Planck collaboration, $\Omega_\chi h^2 = 0.1186\pm 0.0020$~\cite{Aghanim:2018eyx}.
Assuming DM particles are thermally produced in the early Universe, the relic abundance is determined by their thermally averaged annihilation cross sections at the decouple epoch.
If the annihilation cross sections  are too small, DM would be overproduced, contradicting the observation.

The evaluation of DM density is  determined by  the Boltzmann equations.
Assuming standard thermal history of the universe, the DM relic abundance can be parameterized as~\cite{Kolb:1990vq,Jungman:1995df}
\begin{equation}
  \Omega_\chi h^2 \simeq \frac{1.04\times 10^9
  ~\GeV^{-1} (T_0/2.725~\mathrm{K})^3 x_f}{M_\mathrm{pl}\sqrt{g_\star (x_f)}(a+ 3b/x_f)},
\end{equation}
where $x_f \equiv m_\chi /T_f$ with $T_f$ denoting the DM freeze-out temperature.
$g_\star (x_f)$ is the effectively relativistic degrees of freedom at the time of DM freeze-out. $M_\mathrm{pl}$ is the Planck mass and $T_0$ is the present CMB temperature.
$a$ and $b$ are the coefficients in the velocity expansion of annihilation cross section $\sigma_\mathrm{ann} v = a+ b v^2 + \mathcal{O}(v^4)$.
If DM can annihilate into more than one channel, $a$ and $b$ are the total coefficients of all open channels.
In order to calculate relic density for DM involving in this work, we should firstly calculate the $a$ and $b$ coefficients in various annihilation channels.

In the scenario DM interacts with electroweak gauge bosons,  DM can annihilate to $\gamma\gamma$, $ZZ$, $\gamma Z$, and $W^+W^-$.
After EWSB,  $B_\mu$ and $W_\mu^3$ mix into the photon field $A_\mu$ and massive gauge field $Z_\mu$.
The effective interactions in terms of physical fields $A^\mu$ and $Z^\mu$ are
\begin{align}
\mathcal{L} &\supset \frac{C_A}{\Lambda^3} \bar{\chi} \chi F_{\mu\nu}F^{\mu\nu} + \frac{C_{\gamma Z}}{\Lambda^3} \bar{\chi} \chi F_{\mu\nu}Z^{\mu\nu}\nonumber\\
& + \frac{C_{ZZ}}{\Lambda^3} \bar{\chi} \chi Z_{\mu\nu}Z^{\mu\nu}  
 +  \frac{C_{W^+W^-}}{\Lambda^3} \bar{\chi} \chi W_{\mu\nu}W^{\mu\nu} 
\end{align}
with $F_{\mu\nu} = \partial_\mu A_\nu - \partial_\nu A_\mu$, $Z_{\mu\nu} = \partial_\mu Z_\nu - \partial_\nu Z_\mu$,  and $W_{\mu\nu} = \partial_\mu W_\nu^+ - \partial_\nu W_\mu^-$.
 The matching conditions for interactions involving neutral gauge boson are
\begin{align}
C_A &= C_B c_W^2 + C_W s_W^2,   \nonumber\\
C_{\gamma Z} &=  2 s_W c_W (C_W-C_B),\nonumber \\
C_{ZZ} &= C_B s_W^2 + C_W c_W^2.
\end{align}
Interaction with $W$ bosons only origins from $C_W \bar{\chi}\chi W_{\mu\nu}^i W^{i \mu\nu}$, leading to $C_{W^+W^-} = C_W$.

The coefficient $a$ vanishes in each channel, and the leading contribution to $\left<\sigma_\mathrm{ann} v\right>$ is from the $p$-wave.
The coefficients $b$'s for the annihilation channels of $\chi \chi  \to \gamma \gamma$, $\chi \chi  \to ZZ$, $\chi \chi  \to \gamma Z$ and $\chi \chi  \to {W^ + }{W^ - }$ are
\begin{align}
b_{\chi\chi\to \gamma\gamma} &= \frac{{C_{{\mathrm{A}}}^2m_\chi ^4}}{{\pi {\Lambda ^6}}},\nonumber\\
b_{\chi\chi\to ZZ} &= \frac{{C_{{\mathrm{ZZ}}}^2{\rho _Z} m_\chi ^4 (8 - 8 x_Z + 3 x_Z^2)}}{{8\pi {\Lambda ^6}}},\nonumber\\
b_{\chi\chi\to \gamma Z}  &= \frac{C_\mathrm{\gamma Z}^2 \rho_{\gamma Z}^2 (4-x_Z)^2 }{{32\pi {\Lambda ^6}}}, \nonumber\\
b_{\chi \chi  \to {W^ + }{W^ - }}& = \frac{{C_{W^+W^-}^2{\rho _W}m_\chi ^4 (8 - 8 x_W + 3x_W^2)}}{{4\pi {\Lambda ^6}}},
\end{align}
where $x_{Z/W}\equiv m_{Z/W}^2/m_\chi^2$, ${\rho _{Z/W}} \equiv \sqrt {1 - m_{Z/W}^2/m_\chi ^2}$, and $\rho_{\gamma Z} \equiv \sqrt{1- m_Z^2/4m_\chi^2}$.

Summing over all the annihilation channels and writing the coefficients $b$'s in terms of  $C_B$ and $C_W$,  we obtain 
\begin{widetext}
\begin{equation}
\begin{split}
b_B &= \frac{1}{\pi }\frac{{m_\chi ^4}}{{{\Lambda ^6}}}\left( {c_w^4 + s_w^4\frac{{{\rho _Z}(8 - 8{x_Z} + 3x_Z^2)}}{8} + \frac{{c_w^2s_w^2\beta _{\gamma Z}^2{{(4 - {x_Z})}^2}}}{8}} \right) C_B^2, \\
b_W &= \frac{1}{\pi }\frac{{m_\chi ^4}}{{{\Lambda ^6}}}\left( {s_w^4 + c_w^4\frac{{{\rho _Z}(8 - 8{x_Z} + 3x_Z^2)}}{8} + \frac{{c_w^2s_w^2\beta _{\gamma Z}^2{{(4 - {x_Z})}^2}}}{8} + \frac{{{\rho _W}(8 - 8{x_Z} + 3x_Z^2)}}{4}} \right)C_W^2,\\
b_{BW} &= \frac{1}{\pi }\frac{{m_\chi ^4c_w^2s_w^2}}{{{\Lambda ^6}}}\left( {1 + \frac{{{\rho _Z}(8 - 8{x_Z} + 3x_Z^2)}}{4} - \frac{{\beta _{\gamma Z}^2{{(4 - {x_Z})}^2}}}{4}} \right) C_B C_W.
\end{split}
\label{eq:b_eff_ww}
\end{equation}
\end{widetext}
Loop-induced annihilations to quarks and gluons  are subdominant and can be safely ignored. It is known that when $C_W = 0$, the annihilation channel of $\gamma \gamma $ always dominates, and the constraint on $C_B$ is roughly  proportional to the inverse of $m_\chi^2$.
While in the case $C_B = 0$, the annihilation channel of $\gamma Z$ or $W^+W^-$ will dominate once the channel opens, explaining the behavior of relic density curve  with $m_\chi$ around  $80~\GeV$ in Fig.~\ref{fig:eff_ww}(B).

In the case that the DM candidate interacts with leptons, DM decays to leptons directly,  with $a$ and $b$ coefficients given by
\begin{eqnarray}
 a &=& \frac{1}{2\pi \Lambda^4} \sum_l C_l^2 \rho_l (2 m_\chi^2 + m_l^2) \label{eq:eff_lepton_a}, \\
b &=& \frac{1}{2\pi \Lambda^4} \sum_l C_l^2 \rho_l (2 m_\chi^2 + m_l^2) \frac{-4 + 2 m_f^2/m_\chi^2 +  11 m_f^4/m_\chi^4}{24(1-m_f^2/m_\chi^2)(2+m_f^2/m_\chi^2)}.\nonumber
%\label{eq:btotal_leptons}
\end{eqnarray}
Here the sum is performed over all lepton flavors.

\bibliographystyle{apsrev}
\bibliography{effective}
\end{document}